\documentstyle[prd,tighten,aps,epsfig]{revtex}
\begin{document}
\draft

\twocolumn[\hsize\textwidth\columnwidth\hsize\csname
@twocolumnfalse\endcsname
\renewcommand{\theequation}{\thesection . \arabic{equation} }
\title{\bf Classical and Quantum Quintessence Cosmology }

\author{Pedro F. Gonz\'{a}lez-D\'{\i}az}
\address{Centro de F\'{\i}sica ``Miguel Catal\'{a}n'',
Instituto de Matem\'{a}ticas y F\'{\i}sica Fundamental,\\ Consejo
Superior de Investigaciones Cient\'{\i}ficas, Serrano 121, 28006
Madrid (SPAIN)}
\date{June 12, 2001}

\maketitle

\begin{abstract}
This paper implements the idea of considering the instantonic
creation of brane worlds whose five-dimensional bulk contains a
negative cosmological constant and a scalar quintessence field
with time-dependent equation of state, restricting to the case
that the quintessence field couples minimally to
Hilbert-Einstein gravity. We construct an Euclidean formalism,
both for the four- and five-dimensional cases, singling out a
Hamiltonian constraint that depends on the parameter defining
the quintessence state equation. Specializing at several
particular values of that parameter, we obtain solutions to the
constraint equation and analyse them both classically and
quantum mechanically. It is found that these solutions can
represent either asymptotically anti-de Sitter wormholes or pure
anti-de Sitter spaces whose quantum states are obtained by means
of the Wheeler de Witt equation. Starting with the different
five-dimensional solutions, an instantonic procedure is applied
to describe the creation of geometrically equivalent inflating
de Sitter branes whose quantum states are also evaluated in some
cases. We thus consider the quantum state of the universe to be
contributed by all the instantonic paths that correspond to
these particular brane worlds.

\end{abstract}

\vskip2pc]

\renewcommand{\theequation}{\arabic{section}.\arabic{equation}}


\section{Introduction}
\setcounter{equation}{0}

In the opinion of many cosmologists, recent observations have
actually promoted cosmology up to the status of a precision
science. In particular, the measurements made on distant
supernovas Ia [1] and posterior observations and data
refinements [2] have led to the increasingly firmer conclusion
that our universe is now accelerating. On the other hand, the
even more recent baloon-type experiments [3,4] have quite nicely
fixed the spectrum of CMB anisotropies, at least for the initial
region of small angles. Although these observations should be
extended to even larger redshifts and smaller angles, such as
several missions, currently under preparation, will do in the
near future [5,6], it already appears well-proved that around
two-third of the energy in the universe should be hidden as a
dark energy [7]. The most successful and general way to account
for this dark energy has been suggested to be the incorporation
of a cosmological, so-called quintessence field [8] which
advantageously replaces the cosmological constant and can be
nicely fitted to predict both cosmic acceleration and a suitable
spectrum of CMB anisotropies [9].

Quintessence was first conceived [8] as a slowly-varying, scalar
field with constant equation of state, $\omega=Const.<0$ (in
$p=\omega\rho$), but it soon appeared that in order for this
field to allow cosmological scenarios more adaptable to the
foreseeable demands of future observations and more compatible
with possibly related particle physics theories, one had to
endow the quintesence field with the additional degrees of
freedom resulting from a time-variable equation of state and a
larger interval for $\omega$ (i.e. the quintessence field is no
longer necessarily slowly-varying or confined to satisfy a
constant value for $\omega$). Tracking models of quintessence
[10] would moreover allow emergence, after recombination, of an
attractor solution which confortably solves the problem of
cosmic coincidence [11] and may improve, though does not solve,
that of the cosmological constant. In spite of all the successes
that modern quintessence has already achieved, as it happened
with its ancient Greek ancestor, it retains the shortcoming that
one can wonder, where would such a field come from?. A recent
idea that may conveniently answer this question is to consider
[12] that the quintessence field springs from the physics of
extra dimensions, or in other words, that this field should be a
natural component of the higher dimensional bulk of the brane
worlds, and was created when the universe was created from
nothing or "something ". It then appears an apealing idea to
consider the instantonic creation of a brane world whose
five-dimensional bulk contains a negative cosmological constant
and a scalar quintessence field. This paper aims at implementing
this idea within the framework of five-dimensional Euclidean
gravity, restricting ourselves to the case that the quintessence
field minimally couples to Hibert-Einstein gravity.

The physics of the brane worlds created in a bulk with extra
dimensions is other fundamental issue [13] in present cosmology
that may help to solve the long standing hierarchy problem [14].
Our approach in this paper involves an instantonic procedure
[15] which, without resorting to any particular particle-physics
models, will allow us to deal with the process of creation of a
brane world with a quintessence field, using only the machinary
of canonical semiclassical and quantum cosmology. As to the
boundary conditions for a universe created this way, we shall
adopt the general notion of creation from nothing [16], but only
in the sense of allowing also the possibility that the universe
was created from a baby universe that evolved as a component of
the quantum spacetime foam, and actually take these two creative
mechanisms (which e.g. respectively entail a no boundary initial
condition [17] or a regular wave functional as the geometry
degenerates [18]) to be simultaneously contributing the quantum
state of the universe. It can be noted that such a generalized
boundary condition is indeed compatible with current ideas on
the early evolution of a tracking quintessence field in four
dimensions. Thus, as one goes back in cosmological time passing
through the distinct (potential, transition and kinetic)
quintessential regimes, each characterized by given nearly
constant values of $\omega$ [19], one finally arrives at very
high redshifts of the order $z\geq 10^{30}$ where, according to
the classical equations, $\omega$ starts decreasing from
$\omega=+1$. However, at those high values of $z$, the universe
had already actually entered a quantum regime where all possible
values at least from +1 to -1 are allowed for $\omega$.

The paper can be outlined as follows. In Sec. II we describe the
general Euclidean formalism for a homogeneous scalar
quintessence field which is minimally coupled to
Hilbert-Einstein gravity in the presence of a negative
cosmological constant, both in the five- and four-dimensional
cases. From the equations that define the quintessence field and
its conservation law, we finally obtain a Hamiltonian constraint
equation which is expressed in terms of the parameter $\omega$.
Sec. III deals with the classical solution and the quantum state
that correspond to the particular state equation $\omega=-3/4$.
It is shown that the classical solution describes a new
asymptotically anti-de Sitter (AdS) Euclidean wormhole and that,
if we cut the manifold at the wormhole throat, the wave function
describes a pure quantum state which can be given in terms of
parabolic cylinder functions. We also consider that the quantum
state should be given as a mixed density matrix when the
manifold cannot be cutted in two disconnected parts and the
instantonic processes leading to either a single brane or a
string of brane-antibrane pairs, and their respective quantum
states. Other particular classical and quantum Euclidean
solutions corresponding to different values of $\omega$ covering
the entire range of its permissible values are analysed in Sec.
IV, where the role of such solutions in the construction of
brane worlds is also discussed in some detail. Sec. V contains a
general prescription for the cosmological evolution of the
inflating branes after their creation. A tentative model is also
constructed for the evolution of each brane along a cosmological
time in which, starting from the five-dimensional scenario, one
arrives at a general expression relating the scale factor of the
observable universe with the quintessence state equation and the
metrics analysed above on the brane hypersurface. Finally, we
summarize and add some comments on the nucleation probability
for generalized solutions and the graviton ground state and
spectrum of the Kaluza-Klein modes in Sec. VI.

\section{Quintessence dynamics in AdS space}
\setcounter{equation}{0}

In this section we will consider first the general-relativistic
dynamics of a homogeneous quintessential scalar field, $\phi$,
which is minimally coupled to Hilbert-Einstein gravity, in a
five-dimensional framework, in the presence of a negative
cosmological constant $\Lambda_5$. The inclusion of this
negative cosmological constant supports recent claims [20] that
AdS or asymptotically AdS spaces had to have played a rather
decisive role in primordial particle and cosmological physics,
and offers at the same time the opportunity to construct
consistent brane-world scenarios to deal with cosmological and
gravity problems. Originally, quintessence fields were
considered [8] to be slowly varying in order for the pressure
entering their equation of state to be always negative. However,
in more recent quintessence-tracking models [10,19] where that
equation of state is not restricted to be constant, the
quintessence fields are not necessarily assumed to be slowly
varying along the entire cosmological range of redshif values,
but to satisfy a time-variable state equation with the general
form
\begin{equation}
p=\omega(t)\rho ,
\end{equation}
where $p$ and $\rho$ are, respectively, the presure and energy
density for the quintessence field, and $\omega(t)$ is the
time-dependent parameter specifying the equation, with
\begin{equation}
+1 <\omega <-1 .
\end{equation}
As it was pointed out in the Introduction, the time evolution of
parameter $\omega$ in quintessence tracking models is not
specified as one approaches the semiclassical and quantum
regions that correspond to redshifs larger than $\sim
z=10^{30}$, so that one should consider all possible values of
$\omega$, restricted only by Eqn. (2.2), as such regions are
considered. We therefore study a general-relativistic model for
an unspecified equation of state for the quintessence field in
the presence of a negative cosmological constant. Thus, in order
for dealing with the semiclassical and quantum states of our
whole system, instead of choosing the solution for a particular
value of $\omega$, we shall consider a set of such solutions and
interpret them as describing contributing semiclassical or
quantum paths.

The Euclidean action for the five-dimensional system formed by a
homogeneous quintessential scalar field $\phi$ with unspecified
state equation and potential $V(\phi)$, which is minimally
coupled to Hilbert-Einstein gravity in the presence of a
negative cosmological constant $\Lambda_5$, can be written as
\[I= -\frac{1}{16\pi G_5}\int d^5
x\sqrt{g}\left(R+2\Lambda_5\right)\]
\[+\frac{1}{2}\int d^5 x\sqrt{g}\left[\left(\nabla\phi\right)^2
+2V(\phi)\right]\]
\begin{equation}
-\frac{1}{8\pi G_5}\int d^4 x\sqrt{h}{\rm Tr}K ,
\end{equation}
where $G_5$ is the five-dimensional gravitational constant, $g$
and $h$ are the determinant of the five-metric and the metric
induced on the boundary four-surface, respectively, $R$ is the
scalar curvature, and $K$ is the second fundamental form on the
boundary surface. Assuming an ansatz for the Euclidean metric
with topology ${\rm R}\times{\rm S}^4$, i.e.
\begin{equation}
ds^2=N^2 d\tau^2+a(\tau)^2 d\Omega_4^2 ,
\end{equation}
in which $\tau$ is the extra direction and $N$ is the lapse
function, with $d\Omega_4^2$ the metric on the unit four-sphere
and the actual time coordinate for observers in ${\rm S}^4$
being one of the coordinates of that sphere (see Sec. V), the
action (2.3) reduces to
\[I=
-\frac{3}{4\pi G_5}\int d^5 x
Na^2\times\]
\begin{equation}
\left[\frac{\dot{a}^2}{N^2} +1+\Lambda a^2
+\frac{2\pi G_5}{3}\left(\frac{\dot{\phi}^2}{N^2}
+2V(\phi)\right)a^2\right] ,
\end{equation}
where the overhead dot denotes $\tau$-derivative and
$\Lambda=-\Lambda_5/6$. We thus note that even though $\tau$ is
not a true time coordinate for observers in ${\rm S}^4$,
relative to the unobservable five-dimensional space as a whole,
it plays the role of an Euclidean time.

From action (2.5) we can obtain the field equations for the
scale factor $a$ and the field $\phi$ and the Hamiltonian
constraint. In the gauge $N=1$, they are:
\begin{equation}
\frac{d}{d\tau}\left(a\dot{a}\right) =1+\Lambda a^2+\frac{4\pi
G_5 a^2}{3}\left[\dot{\phi}^2+2V(\phi)\right]
\end{equation}
\begin{equation}
\ddot{\phi}+4\dot{\phi}\frac{\dot{a}}{a}=\frac{dV(\phi)}{d\phi}
\end{equation}
\begin{equation}
-\dot{a}^2+1+\Lambda a^2- \frac{2\pi
G_5}{3}\left(-\dot{\phi}^2+2V(\phi)\right)a^2=0 .
\end{equation}
The above expressions can be simplified if we now use the
definition of the five-dimensional quintessence field $\phi$ in
terms of the pressure $p$ and the energy density $\rho$. For
that definition we will simply take the extension of the usual
four-dimensional definition of the quintessence field [8] to the
five-dimensional case, choosing the extra coordinate $\tau$ to
play the role of an Euclidean time. If we rotate this to its
"Lorentzian" counterpart, $\tau\rightarrow -it$, then we have
the following definition for the field $\phi$ in five
dimensions:
\begin{equation}
8\pi G_5\rho =\frac{1}{2}\dot{\phi}^2+V(\phi)
\end{equation}
\begin{equation}
8\pi G_5 p=\frac{1}{2}\dot{\phi}^2-V(\phi) ,
\end{equation}
where the overhead dot means now $d/dt$. As referred to a
four-sphere, we can also introduce a conservation law for field
$\phi$ given by
\begin{equation}
\rho=\rho_0 a^{-4(1+\omega)} ,
\end{equation}
in which $\rho_0$ is a constant. In particular, using Eqns.
(2.9)-(2.11), we obtain a relation between the Euclidean
counterparts of the energy density for the quintessence field
and the scale factor given by
\begin{equation}
-\dot{\phi}^2+2V(\phi)=16\pi G_5 \rho_0 a^{-4(1+\omega)} ,
\end{equation}
with which the Hamiltonian constraint (3.8) reduces to
\begin{equation}
\dot{a}^2-\Lambda a^2+A^2 a^{-2(1+2\omega)}=1 ,
\end{equation}
where $A^2=32\pi^2 G_5^2\rho_0/3$ is a constant. From the state
equation $p=\omega\rho$ and the definition of $\phi$ given by
Eqns. (2.9) and (2.10) it follows that, if we want to preserve
an interval $+1 >\omega > -1$ for $\omega$ also in the Euclidean
region, then the $t$-dependence of $\phi$ and the shape of
potential $V(\phi)$ should be such that the rotation
$t\rightarrow i\tau$ must necessarily entail a change of sign of
potential $V[\phi(t)]\rightarrow -V[\phi(\tau)]$ which satisfies
the equations of motion.

Although we have not been able to find any closed-form solutions
to Eqn. (2.13) for a generic $\omega$, some analytical solutions
for particular values of $\omega$ will be obtained and discussed
in Secs. III and IV. In what follows of the present section, we
shall restrict ourselves to briefly consider the case in which
the coupling between gravity and the quintessence field takes
place in the presence of a negative cosmological constant in
four dimensions. One can readidy show that in that case the
Euclidean action is like Eqn. (2.3), but for a Newton
gravitational constant $G_N$ instead $G_5$, with the integration
being now over four dimensions, i.e. over $d^4 x$ in the main
terms and over $d^3 x$ in the surface term. After applying the
general-relativity machinary for a minisuperspace $ds^2=d\tau^2
+a(\tau)^2 d\Omega_3^2$ (with $d\Omega_3^2$ the metric on the
three-sphere) we obtain then
\[\dot{a}^2-1 -\Lambda a^2\]
\begin{equation}
+\frac{2\pi G_N}{3}\left[-\dot{\phi}^2+2V(\phi)\right]a^2=0 .
\end{equation}
Using the definition of the four-dimensional quintessence field
[8], which is also given by Eqns. (2.9) and (2.10), and a
conservation law referred to a three-surface for it [21],
$\rho=\rho_{40} a^{-3(1+\omega_4)}$, we finally obtain for the
Hamiltonian constraint
\begin{equation}
\dot{a}^2-\Lambda a^2+A_4^2 a^{-(1+3\omega_4)}=1 ,
\end{equation}
where the constant in the third term is now given by
$A_4^2=32\pi^2 G_n^2\rho_{40}/3$. Comparison of expressions
(2.13) and (2.15) leads to the conclusion that all possible
solutions to the constraint (2.15) for a particular equation of
state $\omega_4$ in four dimensions should be the same as the
solutions to Eqn. (2.13) for an equation of state specified by a
parameter $\omega=3\omega_4/4-1/4$ in five dimensions. In the
two following sections we shall explicitly consider some of such
solutions.

\section{The $\omega=-3/4$ five-dimensional wormhole}
\setcounter{equation}{0}

If we set the equation of state for the quintessential field to
be $\omega=-3/4$, then from Eqn. (2.13) we obtain a Lorentzian
Hamiltonian constraint given by
\begin{equation}
-\dot{a}^2-\Lambda a^2 +A^2 a=1 .
\end{equation}
This constraint is equivalent to the equation that describes the
motion of a particle in a potential
\[U(a)=\frac{1}{2}\left(\Lambda a^2-A^2 a+1\right) ,\]
with zero total energy. Potential $U(a)$ has its minimum at
$a_{\rm min}= A^4/(8\Lambda)$. From the constraint equation
(3.1), one can see that an Euclidean solution for time
$\tau\rightarrow -it$ can only exist when $\Lambda >A^4/4$.
Solutions to Eqn. (3.1) which are nonsingular and periodic (in
Lorentzian time) lie in the region $a_-
<a<a_+$, with
\begin{equation}
a_{\pm}=\frac{A^2\pm\sqrt{A^4-4\Lambda}}{2\Lambda} .
\end{equation}
The Euclidean solution will be then
\begin{equation}
a(\tau)=
\frac{A^2\pm\sqrt{\alpha}\sinh\left(\sqrt{\Lambda}\tau\right)}{2\Lambda}
,
\end{equation}
where $\alpha=4\Lambda-A^4 >0$, and the upper + sign accounts
for the region with $\tau>0$ and the lower - sign accounts for
the region with $\tau<0$. This is a new asymptotically anti-de
Sitter Euclidean wormhole solution, which is qualitatively the
same, but formally simpler than the one that corresponds to a
five-dimensional $\omega=0$ state equation [22,27] (see Sec.
IV). As pointed out at the end of Sec. II, a solution which is
formally the same as that is given by Eqn. (3.3) can be obtained
also in the four-dimensional case for an equation of state
$\omega_4=-2/3$, i.e. in the quintessential regime which is able
to reproduce an accelerating universe.

Four-dimensional Euclidean wormholes have always been looked at
with some mistrust because they evolve in Euclidean time and
therefore somehow "live in eternity". Perhaps the physics of
extra dimensions may provide a suitable framework for wormholes
to become more acceptable physically. In fact, five-dimensional
wormhole solutions, such as the one given by Eqn. (3.3), would
evolve along the fifth direction of either a positive definite
metric, or a metric with Lorentzian signature having as time
direction one of the four-sphere (see Sec. V). In the latter
case, wormholes would no longer show the above shortcoming.

In the Lorentzian region, we obtain a real solution only for the
case $A^4 >4\Lambda$, that is
\[a(t)=
\frac{A^2 +\sqrt{-\alpha}\sin\left(\sqrt{\Lambda}t\right)}{2\Lambda},\]
which varies from the radius of the wormhole throat
$a_0=A^2/(2\Lambda)$, at $t=0$, first up to a maximum radius
$\left(A^2+\sqrt{-\alpha}\right)/(2\Lambda)$, then down to
$\left(A^2-\sqrt{-\alpha}\right)/(2\Lambda)$, to finally return
to $a_0$ again, as $t$ completes its period at
$t=2\pi/\sqrt{\Lambda}$. This Lorentzian solution gives the
scale factor of a closed nonsingular baby universe.

Moreover, one can express solution (3.3) in term of the
Euclidean conformal time $\eta$ defined by
\[\eta=\int\frac{d\tau}{a(\tau)}=\]
\begin{equation}
\eta_{*}^{\pm}\pm
2{\rm
arctanh}\left[\frac{A^2\tanh\left(\sqrt{\Lambda}\tau/2\right)
-\sqrt{\alpha}}{2\sqrt{\Lambda}}\right] ,
\end{equation}
where
\begin{equation}
\eta_{*}^{\pm}=\pm 2{\rm
arctanh}\left(\frac{\sqrt{\alpha}}{2\sqrt{\Lambda}}\right),
\end{equation}
with the upper sign + in the above equations accounting for the
conformal time interval $0\leq\eta\leq\eta_{*}^{+} +2{\rm
arctanh}\left[\left(A^2-
\sqrt{\alpha}\right)/\left(2\sqrt{\Lambda}\right)\right]$,
and the lower sign - accounting for $0\leq\eta\leq\eta_{*}^{-}
-2{\rm arctanh}\left[\left(A^2-
\sqrt{\alpha}\right)/\left(2\sqrt{\Lambda}\right)\right]$. In
this way, solution (3.3) can be re-written
\begin{equation}
a(\eta)= \frac{2A^2}{\alpha\cosh\eta-
2\sqrt{\alpha\Lambda}\sinh\eta+A^4} .
\end{equation}

We consider next the quantum state of the above Euclidean
wormhole under the following three usual assumptions: (1) we
disregard all factor ordering problem related with the momentum
operator, (2) the quantum state is restricted to be a pure state
describable by means of a wave function, and (3) we obtain the
wave functional by using the Wheeler de Witt formalism.
Assumption (2) implies that the wormhole manifold is simply
connected in the sense that one can cut it at the wormhole
throat $a_0=A^2/\left(2\Lambda\right)$ to get two disconnected
half-wormhole submanifolds. If the cutting would not divide the
manifold into two disconnected parts, then the quantum state of
the wormhole would be given by a mixed density matrix [23]. One
can also obtain the quantum state by using the Euclidean
path-integral formalism. However, we shall restrict ourselves
here to derive the expression for the wave function by using the
equivalent formalism of the Wheeler de Witt equation. This is
constructed starting with the classical Hamiltonian constraint
(3.1) by replacing the momentum $\pi_a=\dot{a}$, conjugate to
the scale factor $a$, for the quantum-mechanical operator
$1/\left(2\sqrt{\Lambda}\right)\left(\delta/\delta a\right)$. If
we disregard the factor ordering problem, then we can choose for
the wave equation
\begin{equation}
\left[-\frac{1}{4\Lambda}\frac{\partial^2}{\partial a^2}
+\left(\Lambda a^2-A^2 a+1\right)\right]\Psi(a)=0,
\end{equation}
where $\Psi(a)$ is the wave function for each half-wormhole.
Introducing the change of variable
\[z=2\sqrt{\Lambda}a-\frac{A^2}{\sqrt{\Lambda}} ,\]
where the new coordinate $z$ varies from 0, at the wormhole
neck, up to $\infty$ in the asymptotic region, the above wave
equation can be recast in the form
\begin{equation}
\left(-\frac{\partial^2}{\partial z^2} +\frac{1}{4}z^2
+1-\frac{A^4}{4\Lambda}\right)\Psi(z)=0.
\end{equation}
Eqn. (3.8) is a differential equation defining a parabolic
cylinder function $D$ [24,25]. We have then the set of linearly
dependent wave functions characterizing the pure quantum state
of the considered Euclidean asymptotically AdS wormhole:
\begin{equation}
\Psi_1^{\pm}= D_{\frac{A^4}{4\Lambda}-\frac{3}{2}}
\left[\pm\left(2\sqrt{\Lambda}a-\frac{A^2}{\sqrt{\Lambda}}\right)\right]
\end{equation}
\begin{equation}
\Psi_2^{\pm}= D_{\frac{1}{2}-\frac{A^4}{4\Lambda}}
\left[\pm
i\left(2\sqrt{\Lambda}a-\frac{A^2}{\sqrt{\Lambda}}\right)\right].
\end{equation}
Since $\lim_{a\rightarrow\infty}\Psi\rightarrow 0$ and
$\lim_{a\rightarrow 0}\Psi\rightarrow 0$ for all of these wave
functions, they must satisfy the Page-Hawking boundary
conditions for the wave function of Euclidean wormholes [18].

In order to construct brane worlds starting with a
five-dimensional Euclidean metric
\begin{equation}
ds^2= d\tau^2+a(\tau)^2 d\Omega_4^2 ,
\end{equation}
where the scale factor $a(\tau)$ is given by Eqn. (3), we shall
follow the procedure put forwards by Garriga and Sasaki for the
five-dimensional AdS instanton [15]. In case that the manifold
be divisible in disconnected parts by cutting along the radial
extra coordinate, two different situations can be distinguished.
On the one hand, the wormhole manifold is cutted at the neck at
$\tau=0$ and at a given hypersurface $\tau=\tau_0$ (or
$\tau=-\tau_0$), excising the remaining spacetime $\tau <0$ and
$\tau >\tau_0$ (or $\tau>0$ and $\tau <-\tau_0$). Gluing then
the resulting space with a copy of it at the hypersurface
$\tau=\tau_0$ (or $\tau=-\tau_0$), we obtain a single brane if
we introduce a brane tension given by [26]
\begin{equation}
\sigma_{\pm}= \pm\frac{3\sqrt{\alpha\Lambda}
\cosh\left(\sqrt{\Lambda}\tau_0\right)}{4\pi
G_5\left(A^2\pm\sqrt{\alpha}\sinh\left(\sqrt{\Lambda}\tau_0\right)\right)}
,
\end{equation}
where the upper/lower sign stands for $\tau >/< 0$. On the other
hand, if the entire wormhole manifold is cutted at a given
$\tau_0>0$ and at a given $\tau_{0}'<0$, but not at $\tau=0$,
excising the remaining spacetime $\tau >\tau_0$ and
$\tau<\tau_{0}'$, then one can glue a succession of copies of
the resulting spacetime with the $\tau$-orientation reverse of
the respective copies to produce an instanton consisting of an
infinite (or finite if we cut at some $\tau=0$ at right and
left) string of brane-antibrane pairs, each brane with a tension
given by Eqn. (3.12) with the upper sign and each antibrane with
the same tension but with the lower sign. We note that the
brane's tension given in all the above cases by Eq. (3.12)
becomes $\pm 3\sqrt{\alpha\Lambda}/\left(4\pi G_5 A^2\right)$ in
the limit $\tau_0\rightarrow 0$, and approaches the critical
Randall-Sundrum value [14] $\sigma_c=3\sqrt{\Lambda}/\left(4\pi
G_5\right)$ as $\tau_0\rightarrow\infty$. In the limit
$A^2\rightarrow 0$, the tension (3.12) reduces to the singular
expression that corresponds to the case of a pure
five-dimensional AdS space,
$3\sqrt{\Lambda}\coth\left(\sqrt{\Lambda}\tau_0\right)/\left(4\pi
G_5\right)$, such as it happens in the case of the
asymptotically AdS wormhole for $\omega=0$ studied in Ref. [27].

In the first of the above situations, the resulting single-brane
instanton can be illustrated as two opposite Euclidean cups,
each with a Lorentzian baby universe replacing the south pole at
Euclidean time $\tau=0$, which interset each other at
$\tau=\tau_0$. This corresponds to two copies of the spherical
patch of the $\omega=-3/4$ asymptotically AdS half-wormhole bulk
which are bounded by a common four-sphere. Such a four-sphere is
the world-sheet of the single brane. In this way, the
single-brane instanton can be interpreted as a semiclassical
path for the creation of a brane world (containing a bulk
$\omega=-3/4$ Euclidean asymptotically AdS wormhole) from a baby
universe and, when it is cutted in half, the solution
interpolates between the baby universe that replaces the south
pole at $\tau=0$, and a spherical brane of radius
$H^{-1}=a(\tau_0)$ at the equator. As it was pointed out by
Garriga and Sasaki [15] and Bouhmadi and Gonz\'{a}lez-D\'{\i}az [27], a
construction like this is analogous to the four-dimensional de
Sitter instanton [28], though the present instanton describes
creation of a universe from a baby universe, rather than
nothing. In the model describing creation of a string of
brane-antibrane pairs, the whole instanton could be illustrated
as an indefinite chain whose beads are made of two opposite cups
smoothly matching each other at $\tau=0$ (the throat) and
interset the two neighbouring beads at, respectively,
$\tau=\tau_0>0$ and $\tau=\tau_{0}'<0$ (see Fig. 1). The
bounding common four-spheres at $\tau_0 >0$ are the world-sheet
for branes and the bounding common four-spheres at $\tau_{0}'<0$
are the world-sheet for antibranes, and the whole instanton is
to be regarded as a semiclassical path for the creation of a
string of brane-antibrane worlds from a string of baby
universes.

In the case that the four-surface $S$ on which we take the data
does not divide the manifold into two disconnected parts, we
have to use a mixed state density matrix $\rho(a)$ [23], rather
than a wave function $\Psi(a)$. This is usually computed by
using a propagator $K\left[h_{ij},0;h_{ij}',\tau_1\right]$ for
the wave function from the data $[h_{ij}]$ (four-metric) on a
four-surface $S$ at $\tau=0$ to the data $[h_{ij}']$ on another
four-surface $S'$ at $\tau=\tau_1$. This propagator can be
written as
\begin{equation}
K\left[h_{ij},0;h_{ij}',\tau_1\right]= \sum_n
\Psi_{n}\left[h_{ij}\right]\Psi_{n}\left[h_{ij}'\right]
e^{-n\sqrt{\Lambda}\tau_1} ,
\end{equation}

\begin{figure}
\begin{center}
\epsfig{file=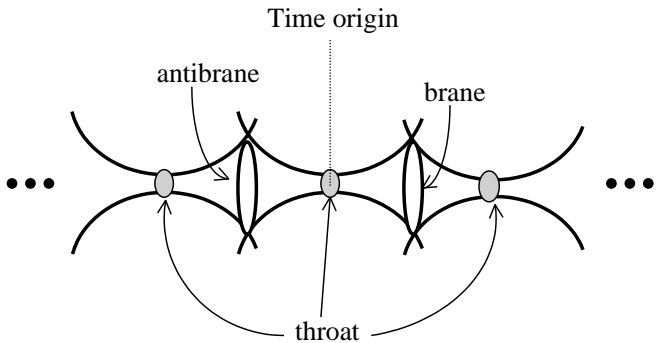,width=\columnwidth}
\end{center}
\caption{Instanton for the creation of a string of brane-antibrane worlds.
Vertical circles represent four-sphere branes or antibranes at
which two $\omega=-3/4$ asymptotically AdS wormholes are glued
introducing a positive or negative tension, respectively. }
\label{rad}
\end{figure}

\noindent where $\Psi_n$ represents the state for the wormhole with an
assumed discrete spectrum labelled by the index $n$. The quantum
state given by a parabolic cylinder function shows nevertheless
a continuous spectrum. However, one could still single out a
discrete spectrum for the wormhole by assuming values for the
parameters $\Lambda$ and $A^2$ such that the index of the
parabolic cylinder function $A^4/4\Lambda-3/4$ in the wave
function (3.10) takes only on positive integer values. Under
such an asuumption, we can re-express the wave function in terms
of Hermite polynomials [24], so that for the upper sign in
solution (3.10) we have
\[\Psi_{n}(n)\propto\]
\begin{equation}
e^{-\frac{1}{2}\left(2\sqrt{\Lambda}a-A^2/\sqrt{\Lambda}\right)^2}
H_{A^4/4\Lambda-3/2}\left(\sqrt{2\Lambda}a-
\frac{A^2}{\sqrt{2\Lambda}}\right) .
\end{equation}
However, if we insist in having positive definite values for the
discrete index, then this wave function cannot be taken to
represent the quantum state in the Euclidean regime, as that
regime is characterized by $4\Lambda > A^4$ for which index $n$
is negative definite. For the Lorentzian regime where $4\Lambda
< A^4$ and the index is positive definite the contribution to the
density matrix from five-geometries that the four-surface $S$
does not divide will be given by the propagator
\begin{equation}
K\left[a,0;a',t_1\right]=
\sum_n^{\infty}\Psi_n[a]\Psi[a']e^{-i(n+1/2)\sqrt{\Lambda}t_1} .
\end{equation}
Because $S$ and $S'$ may have any Lorentzian time separation
$t_1$ in order to obtain the density matrix one has to integrate
over all values of $t_1$ from 0 to $\pi/\sqrt{\Lambda}$. This
gives
\[\rho\left[a;a'\right] =iRe\int_0^{\pi/\sqrt{\Lambda}}dt_1
K\left[a,0;a',t_1\right]\]
\begin{equation}
=2\sum_{n=0}^{\infty}
\frac{\Psi_{n}[a]\Psi_{n}[a']}{(n+1/2)\sqrt{\Lambda}} ,
\end{equation}
where $n=A^4/(4\Lambda)-3/2$. Eqn. (3.16) can be interpreted by
considering that the baby universes associated with the wormhole
are in the quantum state specified by the wave function
$\Psi_{n}[a]$, with a relative probability given by
\[P_n=\frac{2}{(n+1/2)\sqrt{\Lambda}} .\]
Since $A^4 > 4\Lambda$ in the Lorentzian region, the density
matrix given by Eqn. (3.16) is positive definite and can never
diverge.

Finally, the wave function that describes the quantum state of a
single instantonic brane world with the brane at an arbitrary
$a=a_0$ may also be obtained from the single half-wormhole
quantum states given by Eqns. (3.9) and (3.10). Because all of
the spacetime corresponding to $a>a_0$ should here be excised
off, in order to construct the quantum state of a single brane
world, one should integrate the wave functions for the complete
half-wormhole over all values of the scale factor $a$ from the
brane position at $a_0$ to the asymptotic region where
$a\rightarrow\infty$. If we generically represent the wave
function given by expression (3.9) by $\Psi_r=D_r(z)$, with $z$
as given in Eqn. (3.8) and $r=A^4/(4\Lambda)-3/2$, we will then
have for the pure quantum state of a single brane world
\[\Psi_r^{(b)}(z)=\int_{z_0}^{\infty}dzD_r(z)\]
\begin{equation}
=\frac{1}{2}\int_{z_0}^{\infty}dz zD_{r+1}(z)+
\left.D_{r+1}(z)\right|_{z_0}^{\infty} .
\end{equation}
States like this are associated with a continuous spectrum
labelled by the index $r+1$ which is bounded from above at
$r=-1$ due to the Euclidean restriction $4\Lambda > A^4$.
However, if we would assume $r$ to take only on integer values
and then re-express the state (3.17) in terms of Hermite
polynomials [24], it can be readily seen that for $r=-1$ the
quantum state becomes $\Psi_{-1}^{(b)}\propto\exp(-z_0^2/4)$.
Thus, the spectrum associated with the quantum states of a
single brane world constructed starting with a five-dimensional
$\omega=-3/4$ wormhole is generally continuous and runs along
the index $r$ which takes on continuous negative values which
are bounded from above by $r\leq -1$.

\section{Other AdS or asymptotically AdS solutions}
\setcounter{equation}{0}

In this section we shall consider other possible solutions to
the constraint equation (2.13) and discuss their role in the
construction of brane-world instantons similar to those dealt
with in the precedent section. We will study the cases arising
from a quintessential field which is minimally coupled to
Hilbert-Einstein gravity in five and four dimensions. Let us
start with the five-dimensional example for constant equations
of state and the general Euclidean Hamiltonian constraint
\begin{equation}
\dot{a}^2-\Lambda a^2+A^2 a^{-\left(4\omega+2\right)}=1,
\end{equation}
which corresponds to an instantonic metric
\begin{equation}
ds^2 =d\tau^2+a(\tau)^2 d\Omega_4^2 .
\end{equation}
We shall regard the cases corresponding to particular values of
the parameter $\omega$ within the interval $-1 < \omega < +1$.
Thus, for $\omega=+1/4$, we obtain
\begin{equation}
\dot{a}^2-\Lambda a^2+A^2 a^{-3}=1 .
\end{equation}
The Hamiltonian constraint (4.3) also corresponds to the case of
a massless, homogeneous non-quintessential scalar field $\Phi$,
conformally coupled to Hilbert-Einstein gravity in five
dimensions, for which the Euclidean action reads
\[I= - \frac{1}{16\pi G_5}\int d^5 x\sqrt{g}\left[\left(1-
\frac{3\pi G_5 \Phi^2}{2}\right)R +2\Lambda_5\right]
 \]
\begin{equation}
+\frac{1}{2}\int d^5 x\sqrt{g}\left(\nabla\Phi\right)^2
\end{equation}
\[-\frac{1}{8\pi G_5}\int d^4 x\sqrt{h}\left(1- \frac{3\pi
G_5\Phi^2}{2}\right){\rm Tr}K ,\] where $K$ is the extrinsic
curvature on the chosen four-boundary. For an Euclidean metric
of the form
\begin{equation}
ds^2=N^2 d\tau^2+a^2(\tau)d\Omega_4^2 =a(\eta)^2\left(N^2
d\eta^2+d\Omega_4^2\right) ,
\end{equation}
where $N$ again is the lapse function and $\eta=\int d\tau/a$
the conformal time, the Euclidean action becomes \[I=\]
\begin{equation}
-\frac{3}{4\pi G_5}\int d^5 x N\left(a^2\frac{\dot{a}^2}{N^2}
+a^2+\Lambda a^4+\frac{\chi^2}{a}+
\frac{4}{9}a\frac{\dot{\chi}^2}{N^2}\right),
\end{equation}
in which we have redefined $\Lambda=-\Lambda_5/6$ and
$\chi=a^{3/2}\sqrt{3\pi G_5/2}\Phi$. In the gauge $N=1$, we can
derive from this Euclidean action the Hamiltonian constraint and
the equation of motion for the field $\chi$. In terms of the
conformal time $\eta$ these quanties are given by
\begin{equation}
-a'^{2}+a^2+\Lambda a^4+
\frac{1}{a}\left(\chi^2-\frac{4}{9}\chi'^{2}\right)=0
\end{equation}
\begin{equation}
\chi''=\frac{9}{4}\chi .
\end{equation}
A rather trivial solution to Eqn. (4.8) is
\begin{equation}
\chi=A\sinh\left(\frac{3}{2}\right)+B\cosh\left(\frac{3}{2}\right),
\end{equation}
where $A$ and $B$ are arbitrary constants. Hence, we obtain for
the equation of motion for the scale factor,
\begin{equation}
aa''+a'^2-a^2-2\Lambda a^4+aR_0^2=0,
\end{equation}
in which $R_0^2$ is an integration constant, and for the
Hamiltonian constraint the same expression as for the case
$\omega=+1/4$ given by Eqn. (4.3). We have not been able to find
a solution to Eqn. (4.3) in closed form, but only approximate
expressions in limiting cases. Thus, for small $a$, we obtain
$a\simeq A^{2/3}\cosh^{2/3}\left(3\eta/2\right)$, and for large
$a$, the pure AdS solution. This suggests that we have an
asymptotically AdS Euclidean wormhole again for the minimally
coupled quintessence field with state equation $\omega=+1/4$ in
five dimensions. This solution also corresponds to the equation
of state $\omega_4=+2/3$ in four dimensions. Therefore, for the
former case, the cutting and pasting procedure explicited for
the equation of state $\omega=-3/4$ would allow us to construct
essentially the same types of brane-world instantons also in
this case, that is, either a single inflating brane world, or a
string of brane-antibrane worlds. The quantum states for these
wormholes and derived brane worlds will be studied elsewhere.

If we fix the equation of state to describe a pressureless
quintessence field, $\omega=0$, then we obtain from Eqn. (4.1)
\begin{equation}
\dot{a}^2-\Lambda a^2+\frac{A^2}{a^2}=1 .
\end{equation}
The same expression but referred to the constant $A_4$ is also
obtained in the four-dimensional case for a quintessence-field
state equation $\omega_4=+1/3$. It corresponds, on the other
hand, to the case of a homogeneous, massless non-quintessential
scalar field conformally coupled to Hilbert-Einstein gravity in
four dimensions. In terms of the coordinate $\tau$, the solution
of Eqn. (4.11) can be given in closed form and reads:
\begin{equation}
a(\tau)=
\left[\frac{\sqrt{\beta}\cosh\left(2\sqrt{\Lambda}\tau\right)
-1}{2\Lambda}\right]^{1/2} ,
\end{equation}
where
\begin{equation}
\beta=1+4\Lambda A^2 .
\end{equation}
As expressed in terms of the conformal coordinate $\eta=\int
d\tau/a$, solution (4.12) becomes
\begin{equation}
a(\eta)= \sqrt{\frac{\sqrt{\beta}-1}{2\Lambda}}{\rm
nc}\left(\beta^{1/4}\eta\right) ,
\end{equation}
with ${\rm nc}$ a Jacobian elliptic function [24]. This solution
represents an asymptotically AdS Euclidean wormhole which was
first studied by Gonz\'{a}lez-D\'{\i}az [29] and Barcel\'{o} et al. [22],
both in the classical and quantum four-dimensional cases, which
are here associated with an equation of state $\omega_4=+1/3$.
Nevertheless, even though the classical solution (4.12) is
formally the same as that was derived by conformally coupling a
massless scalar field to Euclidean four-dimensional
Hilbert-Einstein gravity [30], the quantum mechanical treatment
given in Ref. [22] no longer applies to the present case because
here the scalar field couples minimally to five-dimensional
gravity and is subjected to the specific definitory
characteristics of a quintessence field constraining it to
depend on $a$ according to Eqn. (2.12). Thus, starting with the
Wheeler de Witt equation derived from Eqn. (4.11) by following
the same procedure as in Sec. III, and redefining the scale
factor so that $x=\sqrt{2\Lambda}a$, we get
\begin{equation}
\left[\frac{\partial^2}{\partial x^2} -\left(x^2-\frac{4\Lambda
A^2}{x^2}+2\right)\right]\Psi(x)=0 .
\end{equation}
Therefore, instead of a function depending on a dicrete index
such as the Hermite polynomials, solutions to the wave equation
can be generally given in terms of the confluent hypergeometric
functions and read
\[\Psi_{\pm}(a)=\]
\begin{equation}
\left(\sqrt{2\Lambda}a\right)^{(1\pm
B)/2}e^{-\Lambda a^2}
\Theta\left(1\pm\frac{1}{4}B,1\pm\frac{1}{2}B,2\Lambda
a^2\right),
\end{equation}
where $B=\sqrt{1-16\Lambda A^2}$ and $\Theta$ represents the
Kummer's functions $M$ and $U$ [24]. The pure quantum states
given by solutions (4.16) are regular as the geometry
degenerates at $a=0$ and exponentially damp as
$a\rightarrow\infty$, so providing satisfactory quantum states
to represent a wormhole. These states will generally span a
continuous spectrum as one lets the parameters $\lambda$ and
$A^2$ to continuously vary. For the limiting case in which
$B=0$, the quantum state would reduce to the wave functions
(4.16) with the confluent hypergeometric function $\Theta$
replaced for the generalized Laguerre polynomial
$L_{-1}^{(0)}(2\Lambda a^2)\equiv L_{-1}(2\Lambda a^2)$, which
identically vanishes.

The use of the instanton (4.12) in the five-dimensional case for
the construction of inflating brane worlds has recently been
made by Bouhmadi and Gonz\'{a}lez-D\'{\i}az [27]. It leads essentially to
the same qualitative picture as that we have described for the
wormhole obtained for a state equation $\omega=-3/4$ in the
five-dimensional case (or $\omega_4=-2/3$ in four dimensions).
Thus, the quantum state of such brane worlds would be obtained
by integrating the quantum states (4.16) over $a$ from $a_0$ to
$a\rightarrow\infty$ . Also in this case there is no quantum
state for the limiting value $B=0$,
\[\Psi^{(b)}=\int da \left(\sqrt{2\Lambda}a\right)^{1/2}
e^{-\Lambda a^2}L_{-1}\left(2\Lambda a^2\right)=0 ,\] so that
there could be no brane world being created in such a limiting
case.

We consider next the situation created in the five-dimensional
framework when $\omega=-1/4$ (or $\omega_4=0$ in four
dimensions). Eqn. (4.1) becomes then
\begin{equation}
\dot{a}^2-\Lambda a^2+\frac{A^2}{a}=1 .
\end{equation}
We have not obtained solutions to this equation in closed form,
but only approximate expressions for the limiting cases. In
fact, whereas for small $a$ we have the wormhole-neck behaviour
\begin{equation}
a\simeq \frac{A^"}{2}\left(1+\sinh\eta\right)\simeq
\frac{A^2}{2}\sqrt{1+\frac{4\tau}{A^2}} ,
\end{equation}
an asymptotic pure AdS behaviour is obtained at large $a$. That
is, although its spacetime structure is quite more complicated
than the solution dealt with in Sec. III, one would expect the
solution to Eqn. (4.17) to also describe an Euclidean
asymptotically AdS wormhole, and to lead to brane-world
instantons having essentially the same characteristics as those
derived for the $\omega=-3/4$ five-dimensional wormhole.
Explicit expressions for the possible quantum states of these
wormholes and brane worlds will be also considered elsewhere.

Finally, let us briefly look at the five-dimensional case
$\omega=-1/2$ (or $\omega_4=-1/3$ in four dimensions). We have
in this case
\begin{equation}
\dot{a}^2-\Lambda a^2+A^2=1 .
\end{equation}
The solution to Eqn. (4.19) is (for $A^2 < 1$):
\begin{equation}
a=
\sqrt{\frac{1-A^2}{\Lambda}}\sinh\left(\sqrt{\Lambda}\tau\right)
=\frac{\sqrt{1-A^2}}{\sqrt{\Lambda}\sinh\left(\sqrt{1-A^2}\eta\right)}
,
\end{equation}
that is, pure AdS space with radius
$r(A)\equiv\sqrt{\frac{1-A^2}{\Lambda}}$. The construction of
brane worlds from Euclidean five-dimensional AdS instanton was
considered by Garriga and Sasaki for the extreme case at
$A^2\rightarrow 0$ [15], i.e. for a AdS radius $r(0)\equiv
1/\sqrt{\Lambda}$. Hence, after conveniently replacing $r(0)$
for $r(A)$, the construction of brane-world instantons here will
follow exactly the same pattern as what was made by these
authors. We thus obtain single brane instantons, but no string
of brane-antibrane instantons can be constructed. On the other
hand, the generalized solution that follows to (4.2) in this
case is singular in the sense that the Weyl tensor diverges as
$\tau\rightarrow 0$, a shortcoming which is not present in all
the generalized solutions that correspond to the asymptotically
AdS wormhole instantons considered in this work.

Let us consider the quantum state that can be defined for our
AdS space. By applying the usual correspondence principle to the
Hamiltonian constraint given by Eqn. (4.19) and disregarding the
factor ordering ambiguity, we derive the following Wheeler de
Witt wave equation:
\begin{equation}
\left[\frac{1}{4\Lambda}\frac{\partial^2}{\partial a^2}
+\Lambda a^2+\left(1-A^2\right)\right]\Psi^{(AdS)}(a)=0.
\end{equation}
Inserting the complex coordinate transformation
\[a=\frac{1}{2\sqrt{\Lambda}}e^{i\pi/4}b\]
and
\[1-A^2\rightarrow i\left(1-A^2\right) ,\]
in this equation, we obtain
\begin{equation}
\left(\frac{\partial^2}{\partial b^2} -\frac{1}{4}b^2
+\left(A^2-1\right)\right)\Psi^{(AdS)}(b)=0 .
\end{equation}
Lineraly dependent solutions to Eqn. (4.22) can also be given in
terms of the parabolic cylinder functions $D$. They are [25]
\begin{equation}
\Psi^{(AdS)}_{\pm}(a)= D_{i\left(A^2-1\right)-1/2}\left(\pm
2\sqrt{\Lambda}ae^{-i\pi/4}\right) ,
\end{equation}
and their complex conjugate counter-parts
$\left(\Psi^{(AdS)}_{\pm}(a)\right)^{*}$. By using the Airy-function
representation of the parabolic cylindric functions [24,25], it
can be seen that all these wave functions satisfy the reasonable
boundary conditions $\lim
\Psi_{a\rightarrow 0}\rightarrow 0$, and $\lim \Psi_{a\rightarrow
\infty}\rightarrow 0$.

Following the procedure leading to Eqn. (3.17), we can now
construct the pure quantum state that corresponds to the single
AdS brane-world instanton. Thus, denoting
$z=2\sqrt{\Lambda}a\exp(-i\pi/4)$, we obtain from Eqn. (4.23)
the state
\[\Psi^{(AdS)(b)}(z)\]
\begin{equation}
=\frac{1}{2}\int_{z_0}^{\infty}dz zD_{i(A^2-1)+1/2}(z)+
\left.D_{i(A^2-1)+1/2}(z)\right|_{z_0}^{\infty} .
\end{equation}
If we now allow $A^2$ to be a complex quantity, $A^2=1-iA_c^2$,
and let $A_c^2$ to continuously vary from 0 to $\infty$, we
would again obtain a continuous spectrum with the ground state
at $A_c^2=1/2$ given by $\Psi_{0}^{(AdS)(b)}\propto\exp(i\Lambda
a_0^2)$.

\section{Primordial cosmological evolution}
\setcounter{equation}{0}

The evolution of the branes after creation would follow the same
general pattern for all particular solutions (all particular
five-dimensional equations of state) considered in the precedent
two sections. It will always be given by analytically continuing
metric (4.2) to real time. Expliciting the metric on the unit
four-sphere as $d\Omega_4^2=d\xi^2+\sin^2\xi d\Omega_3^2$, where
$d\Omega_3^2$ is the metric on the unit three-sphere, we take
$\xi$ as the coordinate playing the role of imaginary time
(recall that $\tau$ is a radial extra coordinate in the
five-dimensional manifold). Assuming for the analytical
continuation $\xi\rightarrow iHt+\pi/2$, with $H=1/a(\tau_0)$,
metric (4.2) becomes
\begin{equation}
ds^2= d\tau^2+a(\tau)^2\left[-H^2 dt^2
+\cosh^2\left(Ht\right)d\Omega_3^2\right] .
\end{equation}
Every contributing brane will therefore inflate with a
characteristic Hubble rate $H$ which distinguishes solutions
from each other. On the brane hypersurfaces at $\tau=\tau_0$,
metric (5.1) reduces exactly to the line element of a
four-dimensional de Sitter metric for each of the different
Hubble rates. Metric (5.1) does not cover the whole spacetime
but only the region from the brane position to the horizon at
$\tau=0$. The remaining region beyond $\tau=0$, which describes
the maximal extension of the corresponding baby universe
spacetime [27,31], is obtained by analytically continuing
$\tau\rightarrow ir$ and $Ht\rightarrow H\varphi-i\pi/2$, that
is
\begin{equation}
ds^2= dr^2-a(r)^2\left[H^2 d\varphi^2
+\sinh^2\left(H\varphi\right)d\Omega_3^2\right] .
\end{equation}

The difference in Hubble rate of the distinct contributing
solutions amounts to the feature that such solutions can be
distinguished from each other by the nature of the inflationary
process driven on them. It is worth noticing by instance that
for the five-dimensional case with $\omega=-3/4$ it can be
obtained that inflation will occur on the brane with a Hubble
parameter given by
\begin{equation}
H_{\pm}= \frac{1}{2}\left[A^2 \pm \sqrt{A^4-4\Lambda\left(1
-\frac{\sigma_{\pm}^2}{\sigma_c^2}\right)}\right] ,
\end{equation}
with the upper sign + for the brane at $\tau>0$ and the lower
sign - for the antibrane at $\tau<0$. This parameter becomes
that of the purely AdS space in the limit $A^2\rightarrow 0$.
Assuming that there is no dilatonic scalar field with nonzero
effective potential [32], if $|\sigma_{\pm}|\geq\sigma_c$, we obtain
that there will be inflation only on the brane, but not on the
antibrane. On the contray, if $|\sigma_{\pm}|<\sigma_c$, then there
will be inflation both on the brane and the antibrane.

On the other hand, once the brane (or antibrane) hypersurface is
fixed, all the initially contributing five-dimensional
instantons considered in this work describe creation of a de
Sitter space and can therefore be regarded to be contributing
semiclassical paths for the creation of the universe. Now, what
about the creation of a four-dimensional quintessence field in
the brane?. Since the five-dimensional quintessence field $\phi$
was assumed to depend only on the extra coordinate $\tau$ (or
$r$), the evolution of that field in the bulk becomes frozen on
the brane with metric
$ds^2=-dt^2+\cosh^2\left(Ht\right)d\Omega_3^2$. We contend that
there may be essentially two ways for the creation of
four-dimensional quintessence fields in the inflating branes.
Firstly, one may resort to cosmological perturbations on the
three-sphere $\Omega_{ij}$ starting from the above metric, so
that $\Omega_{ij}\rightarrow\Omega_{ij}+\epsilon_{ij}$, with the
perturbations $\epsilon_{ij}$ being expanded in terms of the
usual tensor, vector (pure gauge) and scalar harmonics, and the
scalar field $\phi$ being perturbed so that
$\phi\rightarrow\varphi(t)=\phi(\tau_0)+\delta(t)$, in which
$\phi(\tau_0)$ is the frozen constant and $\delta(t)=\sum_n
f_n(t)Q^{(n)}$, with the coefficients $f_n(t)$ depending on the
Lorentzian time entering the four-dimensional metric and
$Q^{(n)}$ the scalar harmonics (note that $n$ simply denotes the
set of labels $n,l,m$). If we next assume the perturbed field
$\varphi(t)$ to satisfy a definition which is given by the
four-dimensional counterpart of Eqns. (2.9) and (2.10) for the
potential $V[\varphi(t)]$, and a conservation law
$\rho_{\varphi}=\rho_{\varphi
0}R^{-3\left(1+\omega_{\varphi}(t)\right)}$, with $R$ the scale
factor and $\omega_{\varphi}(t)=p_{\varphi}/\rho_{\varphi} >-1$,
then we had created an usual four-dimensional quintessence
tracking scenario on the brane world from our original model. On
the other hand, besides other brane models in which the extra
radial coordinate decreases from a very large initial value
during cosmological evolution [33], it is also possible to
envisage an alternative cosmological model derived from the
results of the present work where the coordinate $\tau_0$ on the
brane (or antibrane) hypersurface increases from a minimal small
initial value to follow the same pattern of cosmological
expansion as the observable dimensions. In fact, all the de
Sitter universes which are created through semiclassical paths
fixed by the different possible values of the state equation
parameter $\omega$ in five dimensions can be characterized by a
generic cosmological constant given by
\[\Lambda=\frac{3}{a(\tau_0)^2} ,\]
where $a(\tau_0)$ is the constant value of the scale factor for
each solution at the given brane hypersurface. If we would allow
$\tau_0$ to be a variable, then the cosmological constant will
be variable too and could be looked at like though it were
originated by a cosmic dark field, or "effective" quintessence
field $\bar{\phi}$ in four-dimensional space, with equation of
state $\rho_{\bar{\phi}}=\bar{\omega}p_{\bar{\phi}}$ and energy
density given by
\begin{equation}
\rho_{\bar{\phi}}=\rho_{\bar{\phi}_0}\left(\frac{R_0}{R}\right)^{3\left(1+
\bar{\omega}\right)}= \frac{\Gamma_0}{a(\tau_0)^2} ,
\end{equation}
where $\Gamma_0$ is a constant. We then have
\begin{equation}
\frac{R_0}{R}= \left(\frac{\Gamma_0}{a(\tau_0)^2}\right)^{1/\left[3(1
+\bar{\omega})\right]}.
\end{equation}
On the other hand, one can also fix a general expression for the
potential of the so-generated four-dimensional quintessence
field $\bar{\phi}$ to be
\begin{equation}
V(\bar{\phi})=V_0\left(\frac{R_0}{R}\right)^{3(1+ \bar{\omega})}
=
\frac{\Gamma_0}{a(\tau_0)^2},
\end{equation}
where $V_0$ is a constant. It follows that the four-dimensional
"effective" quintessence field must be proportional to $\tau_0$.
Thus, for the purely AdS path in the regime where $\tau_0 << 1$,
we have $V\propto 1/(\bar{\phi}-\bar{\phi}_0)^2$, which is an
inverse power-law potential whose shape has already been
considered in the literature [34]. Similar limiting expression
for the potential are obtained also for the other considered
solutions.

The above tentative "effective" quintessence model can be
considered as a possible realization of the idea that the
cosmological quintessence field springs from the physics of
extra dimensions [12]. In the present case, the initial
five-dimensional quintessence field can, in turn, be regarded as
the result of the application of a similar mechanism to the
extra sixth coordinate of a six-dimensional space minimally
coupled to a six-dimensional quintessence field, and so on.

\section{Conclusions and further comments}
\setcounter{equation}{0}

This paper contains a classical and quantum treatment of a
primordial five-dimensional Friedmann-Robertson-Walker spacetime
which is endowed with a negative cosmological constant and a
quintessence field with variable state equation. The former is
thought of as being responsible for the connection with particle
physics and the latter is assumed to account for the content of
dark energy needed to justify present cosmological observations.
In this framework, the observable four-dimensional universe
results in the form of a brane world by using an instantonic
procedure.

After briefly analysing current tracking models in the early
cosmological evolution, an Euclidean formalism is developed in
which five- and four-dimensional gravity are minimally coupled
to a scalar, homogeneous quintessence field, in the presence of
a negative cosmological constant. We have studied the possible
interrelations among the obtained solutions to the field
equations and constraints for different state equations of the
quintessence field, both in the four- and five-dimensional
frameworks, and found that such solutions belong to only two
broad categories: they represent either asymptotically AdS
wormhole spaces or pure AdS spaces. In this way, we have
uncovered new asymptotically AdS wormhole spacetimes. Following
then a cutting a pasting procedure, it has been possible to
construct consistent brane worlds from the given instantonic
solutions. We have seen that whenever a five-dimensional
instantonic solution describes a wormhole, one can construct an
infinite or finite string of branes or brane-antibrane pairs. At
least in the cases where it is possible to get solutions in
closed form, one can also obtain explicit expressions for the
quantum state of both, the AdS and asymptotically AdS wormhole
spaces, so as for the brane worlds constructed from these
spacetimes.

Our study is based on the idea that the universe contains a
tracking quintessence field component whose state equation,
$\omega$, may take on any value, only restricted by $-1 <\omega<
+1$, during its very early quantum evolution. Therefore, the
quantum state of the very early universe should be contributed
by the semiclassical or quantum paths that correspond to the
Euclidean solutions associated with all possible values of
$\omega$. As these solutions grow up, they will approach a
spacetime structure whose observable four-dimensional sections
exactly match a de Sitter space and can give rise, thereby, to
fully equivalent four-dimensional geometric structures of the
resulting inflating brane worlds, even though their internal
bulk are different. This may be regarded as a geometrical no
hair theorem for the creation of the universe: no matter the
initial conditions, the universe is always created as an
inflating de Sitter space.

The contribution of each particular solution for a given value
of $\omega$ can be estimated by using the semiclassical
expression $\Gamma\propto\left(-S_E\right)$, where $S_E$ may be
taken to be the Euclidean action that corresponds to the
nonlinear generalization of the metric on the brane, $ds^2
=d\tau^2+a(\tau)^2\gamma_{\mu\nu}dx^{\mu}dx^{\nu}$, where
$\gamma_{\mu\nu}$ is the four-dimensional metric and $a(\tau$ is
the given classical solution. The Euclidean action for
$\omega=0$ and $\omega=-1/2$ were calculated in Refs. [27] and
[15], respectively, for the general case of two concentrical
branes fixed at particular values of the extra dimension [15],
either both at $\tau >0$ or both at $\tau <0$. None of them is
diverging for finite values of the brane positions. For the case
of the generalized solution that correspond to two single
concentrical branes at $\tau_0$ and $\tau_1$ with $\omega=-3/4$,
we obtain
\[S_E=
\sum_{i=0}^{1}(-1)^i \left\{F\tau_i
+\cosh\left(\sqrt{\Lambda}\tau_i\right)
\left[G\sinh\left(\sqrt{\Lambda}\tau_i\right)\right.\right.\]
\[\left.\left.+H\sinh^2\left(\sqrt{\Lambda}\tau_i\right)+
I\sinh^3\left(\sqrt{\Lambda}\tau_i\right)
+J\right]\right\}S_E^{(i)} ,\]
where
\[F=\frac{A^8+
\frac{3}{8}\alpha^2-3A^4\alpha}{4\Lambda^{3/2}a_i^2}\]
\[G= \frac{3\left(2A^4-\alpha/8\right)}{4\Lambda^2 a_i^2}\]
\[H=\frac{13A^2\sqrt{\alpha}}{12\Lambda^2 a_i^2}\]
\[I=\frac{5\alpha}{16\Lambda^2 a_i^2}\]
\[J=\frac{A^2\left(5A^4-8\alpha/3\right)}{8\Lambda^3 a_i^2} ,\]
with $a_i\equiv a(\tau_i)$ and
\[S_E^{(i)}= -\frac{V_4^{(\gamma)}}{4\pi G_N H_i^2} ,\]
in which $V_4^{(\gamma)}$ is the dimensionless volume of the
manifold with metric $\gamma_{\mu\nu}$, $G_N
=G_5\Lambda^{3/2}/\alpha$ is the Newton constant and $H_i
=a_i^{-1}$ is the Hubble constant at the brane at $\tau_i$.
Again this action gives nonvanishing contributions for finite
values of $\tau_i$. Following the procedure described in Ref.
[27], one can also compute the nucleation probability for a
finite or infinite string of brane-antibrane pairs for solutions
which represent asymptotically AdS wormholes. This gives
convengent nonzero contributions for all cases, provided $S_E$
is nonvanishing. Such solutions are, moreover, nonsingular as
the Weyl tensor constructed from them is always finite, even for
single branes, a result which is no longer valid for the case
that the solution describes a pure AdS spacetime [15].

On the other hand, all the solutions considered in this paper
can be related with a ground state for gravity on the
corresponding brane. This state and the corresponding
Kaluza-Klein spectrum should be obtained by solving the
equations of motion for the metric perturbations expressed in
terms of the conformal time [14,15]. In all the cases, the
normalizable ground state corresponding to the massless graviton
is proportional to $a^{3/2}$ and the spectrum of Kaluza-Klein
excitations is continuous and separated by a gap from the zero
mode [15]. We shall consider these issues quite more in detail
in a future publication.

\acknowledgements

The author thanks A.I. Zhuk and M. Bouhmadi for enlightening
discussions and the checking of some calculations and C.L.
Sig\"{u}enza for useful comments and a careful reading of the
manuscript. This work was supported by DGICYT under Research
Project No. PB97-1218.

\end{document}